\documentclass[11pt,aps,pra,reprint,twocolumn,showpacs,superscriptaddress]{revtex4-2}

\usepackage[colorlinks=true,linkcolor=blue,citecolor=blue,urlcolor=magenta]{hyperref}

\usepackage{amsfonts,mathrsfs,amsmath,amsthm,amssymb}
\usepackage{epsfig}
\usepackage{bm}
\usepackage{array,multirow}
\usepackage{booktabs}
\usepackage{graphicx}
\usepackage{upgreek}
\usepackage{xcolor}

\begin{document}  
\title { 
Interference-induced state engineering and Hamiltonian control for noisy collective-spin metrology}

\author{Le Bin Ho} 
\thanks{Electronic address: binho@fris.tohoku.ac.jp}
\affiliation{Department of Applied Physics, 
Graduate School of Engineering, 
Tohoku University, 
Sendai 980-8579, Japan}
\affiliation{Frontier Research Institute 
for Interdisciplinary Sciences, 
Tohoku University, Sendai 980-8578, Japan}

\author{Vu Xuan Tung Duong} 
\affiliation{Department of Mechanical and Aerospace Engineering, Sendai 980-0845, Japan}

\author{Nozomu Takahashi} 
\affiliation{Department of Applied Physics,
Graduate School of Engineering, 
Tohoku University, 
Sendai 980-8579, Japan}

\author{Hiroaki Matsueda}
\affiliation{Department of Applied Physics,
Graduate School of Engineering, 
Tohoku University, 
Sendai 980-8579, Japan}
\affiliation{Center for Science and Innovation in Spintronics, Tohoku University, Sendai 980-8577, Japan}

\date{\today}

\begin{abstract}
Interference provides a fundamental mechanism for generating and manipulating entanglement in many-body quantum systems. Here, we develop an interference framework in which the nonlinear dynamics of collective spin-$\tfrac{1}{2}$ ensembles are mapped onto phase accumulation and self-interference in phase space, providing a direct and physically transparent description of entanglement formation. Within this framework, one-axis twisting produces Greenberger-Horne-Zeilinger (GHZ) states, while two-axis twisting generates multi-component GHZ superpositions relevant for multiparameter quantum metrology. Building on this interference-based description, we analyze metrological performance under realistic Markovian noise, including local and collective emission, pumping, and dephasing, and examine the role of Hamiltonian control based on linear and nonlinear interactions. We show that the optimal control enhances sensitivity in both single- and multiparameter estimation across noise-dependent regimes. These results establish interference as a unifying principle linking nonlinear dynamics, entanglement generation, and metrological performance. This framework offers a broadly applicable route to robust quantum-enhanced sensing in noisy many-body systems.
\end{abstract}

\maketitle

\sloppy

\section{Introduction}
Quantum metrology exploits nonclassical resources, such as entanglement and squeezing, to enhance measurement precision beyond what is achievable using independent subsystems \cite{PhysRevLett.96.010401,RevModPhys.89.035002}. For a probe consisting of \(N\) independent subsystems, the achievable precision is bounded by the standard quantum limit (SQL), scaling as \(1/\sqrt{N}\). By introducing quantum correlations among the subsystems, this bound can be surpassed and, in ideal conditions, the precision can approach the Heisenberg limit with scaling \(1/N\). Such quantum-enhanced sensitivities have enabled substantial improvements in precision technologies including atomic clocks~\cite{RevModPhys.83.331, RevModPhys.87.637,Pedrozo2020,Robinson2024}, magnetometers~\cite{Taylor2008,Nguyen2024,doi:10.1126/science.aam5532}, gravitational-wave detectors~\cite{Schnabel2010,Liu2020,doi:10.1126/science.ado8069,Tobar2024}, dark-matter detection~\cite{PhysRevA.111.012601,Jiang2024,rv43-54zq,cwx5-2n1y}, and quantum-enhanced imaging~\cite{PhysRevLett.133.093601,Xie2025,Pearce2026}.

Collective spin-$\frac{1}{2}$ ensembles provide an attractive and experimentally relevant platform for quantum metrology due to their high controllability and naturally collective dynamics~\cite{PhysRevA.78.052101,PhysRevA.81.021804,PhysRevLett.127.093602,Poggi2024measurementinduced,VIET2023108686,Zheng2022,PhysRevLett.123.260505}. The coherent spin state (CSS) serves as a classical reference probe but is restricted to SQL performance~\cite{MA201189}. Quantum-enhanced metrology therefore requires generating nonclassical correlations, typically via nonlinear collective interactions that produce squeezed or highly entangled many-body states~\cite{MA201189}.

Nonlinear collective interactions provide a powerful mechanism for generating nonclassical collective-spin states for quantum-enhanced sensing. A paradigmatic example is one-axis twisting (OAT)~\cite{PhysRevA.46.R6797,PhysRevA.47.5138,RevModPhys.90.035005,10.1063/5.0204102}, which can generate strong spin squeezing~\cite{PhysRevLett.129.250402,PhysRevA.87.051801,PhysRevApplied.17.064050,ryvz-fksg} and, at specific evolution times, produce maximally entangled Greenberger-Horne-Zeilinger (GHZ) states with optimal noiseless sensitivity~\cite{PhysRevLett.90.030402,PhysRevB.94.205118,PhysRevLett.132.113402}. To achieve faster and more symmetric entanglement generation, two-axis twisting (TAT), has also been proposed and can produce more complex GHZ superpositions relevant for multiparameter metrology~\cite{PhysRevResearch.6.033292}. 

Beyond state preparation, nonlinear collective dynamics further enable interaction-based quantum metrology, where entanglement generation, parameter encoding, and echo-type readout can be interleaved within a single evolution, yielding improved robustness against detection noise and dephasing \cite{PhysRevA.94.010102,PhysRevLett.124.060402,PhysRevResearch.5.043279,Schulte2020ramsey,PhysRevLett.116.053601,PhysRevLett.119.193601,PhysRevA.98.012129}. 
In particular, OAT dynamics can also serve as an interaction-based readout that maps phase information onto collective population observables, enabling Heisenberg-limited sensitivity without single-particle-resolved detection~\cite{PhysRevA.98.012129}.
Related strategies have been explored across OAT, TAT, and twist-and-turn (TNT) dynamics \cite{PhysRevA.98.030303,PhysRevA.97.053618,PhysRevA.97.043813} and supported by experiments demonstrating substantial sensing enhancement in collective spin systems \cite{Chalopin2018}, reinforcing the perspective that Hamiltonian engineering can enable robust quantum metrology beyond maximally entangled probes \cite{RevModPhys.90.035006,PhysRevResearch.6.033292}.

In realistic sensing platforms, decoherence remains a fundamental limitation. Spin ensembles are subject to both local and collective noise processes~\cite{PhysRevA.98.063815}, including emission, pumping, and dephasing, which degrade quantum coherence and suppress the quantum Fisher information (QFI), thereby imposing an optimal sensing time beyond which the estimation precision rapidly degrades. A broad range of control strategies has been developed to counteract these effects. Optimization-based approaches directly target metrological performance by maximizing the QFI under open-system dynamics, using variational formulations~\cite{PhysRevLett.128.160505,PhysRevA.110.052615}, adaptive feedback~\cite{Pang2017,PhysRevA.103.052607}, 
optimal control~\cite{Titum2021,6yzb-43rs}, and hybrid quantum-classical algorithms to co-design state preparation, parameter encoding, and measurement~\cite{Yang2021,Meyer2021,Le2023,MacLellan2024}. At the physical implementation level, Hamiltonian engineering and time-dependent modulation reshape the effective system-environment coupling where dynamical decoupling and related pulse sequences average out unwanted interactions, suppress dephasing, and stabilize QFI scaling~\cite{PhysRevLett.131.220803,Sekatski_2016,JI2021165689,lahcen2025restoringheisenberglimitedprecisionnonmarkovian}. In addition, interaction-based echo protocols exploit controlled nonlinear dynamics to reverse phase diffusion and partially recover metrological gain even in the presence of dissipation~\cite{Riedel2010,PhysRevLett.119.193601}. Together, these developments establish quantum control as a central tool for extending the useful sensing window and preserving quantum-enhanced precision in noisy many-body systems.

In this work, we develop an interference-based framework for collective-spin dynamics and employ it to assess quantum metrology under realistic decoherence.  We begin by analyzing state formation under the OAT and TAT dynamics, showing that the emergence of GHZ and multi-component GHZ states can be understood as self-interference in phase space, which establishes the metrological relevance of the generated entanglement. We next evaluate how realistic noise degrades this entanglement and suppresses the QFI, identifying the regimes in which quantum-enhanced precision survives decoherence. We then investigate Hamiltonian control protocols based on linear and nonlinear interactions, showing how control mitigates noise-induced degradation for both single- and multiparameter estimation. This work thus provides a complete and self-consistent picture of quantum sensing with spin ensembles under realistic noise, from state generation to noise resilience, and identifies the conditions under which collective control yields genuine metrological advantage.

The remainder of this work is organized as follows. In Sec.~\ref{sec2}, we present the interference-based framework and analyze the generation of GHZ and multi-component GHZ states under nonlinear collective-spin dynamics. Sec.~\ref{sec3} investigates single-parameter metrology with GHZ probes in the presence of realistic noise. In Sec.~\ref{sec4}, we examine control-assisted metrology, focusing on Hamiltonian engineering and its impact on noise resilience. Sec.~\ref{sec5} extends the analysis to multiparameter magnetic-field estimation using multi-component GHZ probes. 
Finally, Sec.~\ref{sec6} summarizes the main results and outlines future directions.

\begin{figure*}[t]
    \centering
    \includegraphics[width=0.8\linewidth]{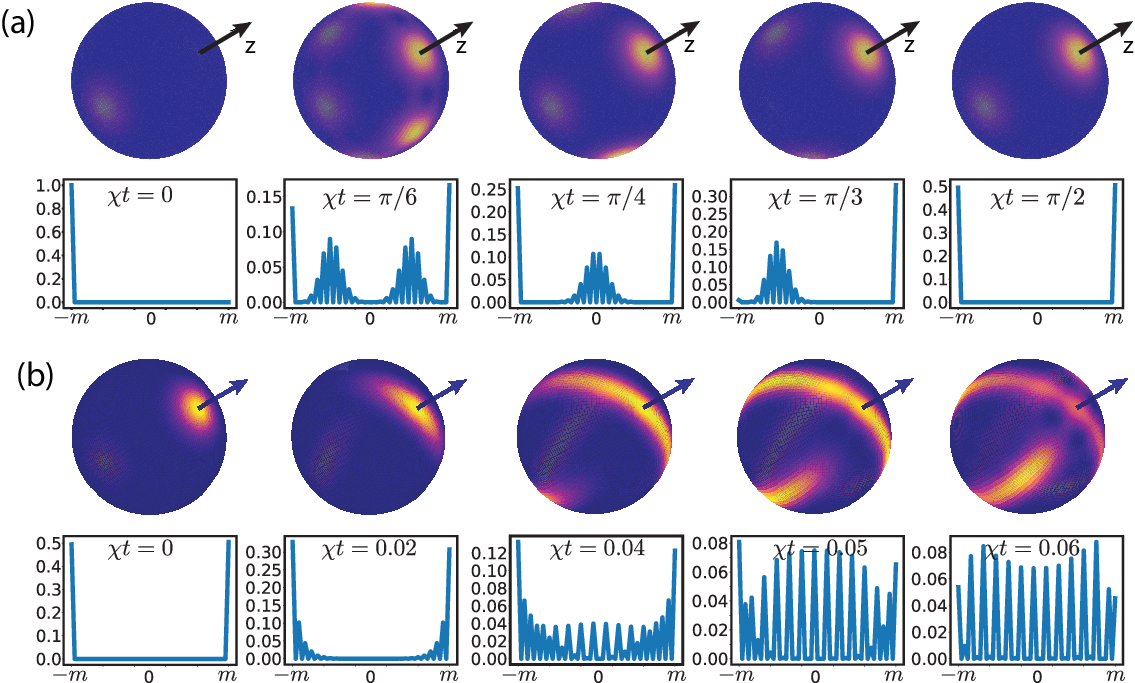}
    \caption{
    (a) Evolution of a spin ensemble under OAT. Starting from the coherent state $|N/2, -N/2\rangle$, the system undergoes squeezing and self-interference, producing constructive and destructive interference patterns in the $z$-$y$ plane. At $\chi t = \pi/2$, this leads to the formation of a GHZ state, $|\psi_{\mathrm{GHZ}_z}\rangle = \frac{1}{\sqrt{2}} \left( |N/2,N/2\rangle + |N/2, -N/2\rangle \right)$.
    (b) Evolution under TAT, starting from the GHZ state $|\psi_{\mathrm{GHZ}_z}\rangle$. The state undergoes squeezing along multiple directions, gradually developing into a multi-component GHZ state. Results are shown for $N = 50$.}
    \label{fig1}
\end{figure*}

\section{Interference-based generation of GHZ and multi-GHZ-like states}\label{sec2}
\subsection{Spin ensemble}
We consider an ensemble of $N$ spin-$\frac{1}{2}$ particles with the joint Hilbert space
\(
\mathcal{H}=(\mathbb{C}^2)^{\otimes N}
\). The collective behavior of the ensemble is captured by the total spin operators
\(
J_\mu=\frac{1}{2}\sum_{i=1}^N\sigma_\mu^{(i)}, \ \mu\in\{x,y,z\},
\)
where $\sigma_\mu^{(i)}$ is the Pauli matrix acting on the $i$th spin. These operators generate a collective representation of the rotation group $\mathrm{SU}(2)$ on $\mathcal{H}$ and satisfy the angular-momentum algebra
\(
[J_\mu,J_\nu]=i\,\epsilon_{\mu\nu\pi}\,J_\pi,
\)
with $\epsilon_{\mu\nu\pi}$ denotes the Levi-Civita symbol. 
The representation $(\tfrac{1}{2})^{\otimes N}$ can be decomposed into irreducible subspaces labeled by the total spin quantum number $j$,
\begin{align}
(\tfrac{1}{2})^{\otimes N}
=
\bigoplus_{j = \{0 \text{ or } 1/2\}}^{N/2}\,\mathcal{H}_j\otimes\mathbb{C}^{d_j},
\end{align}
where $\mathcal{H}_j$ is the spin-$j$ subspace of dimension $2j+1$, and $d_j$ is its multiplicity \cite{SakuraiNapolitano2010}.
The standard basis in each sector is given by 
the Dicke state $|j,m\rangle$, which is the eigenstate of $J^2$ and $J_z$,
\begin{align}
J^2|j,m\rangle=j(j+1)|j,m\rangle,\
J_z|j,m\rangle=m|j,m\rangle,
\end{align}
with $m=-j,-j+1,\dots,j$. The largest sector $j=N/2$ corresponds to the fully symmetric subspace, which is invariant under particle permutations and has the reduced dimension $N+1$ \cite{French_Rickles_2003}.

In many collective-spin platforms, the dynamics preserve permutation symmetry \cite{PhysRevA.78.015804,Hansenne2022,PhysRevA.88.012305,PhysRevD.106.106020,Ho_2019} and therefore remain in the symmetric sector $j=N/2$. In the presence of decoherence or imperfections that break permutation symmetry, the system can populate sectors with total spin $j<N/2$ \cite{PhysRevA.102.022602}, and the dynamics are no longer confined to the symmetric manifold. In the fully symmetric sector $j=N/2$, these states admit the explicit form
\begin{align}
\Big|\frac{N}{2},m\Big\rangle
=
\frac{1}{\sqrt{\binom{N}{\frac{N}{2}+m}}}
\sum_{\text{permutations}}
|\uparrow\cdots\uparrow\downarrow\cdots\downarrow\rangle,
\end{align}
where $\frac{N}{2}+m$ spins are in $|\!\!\uparrow\rangle$ and $\frac{N}{2}-m$ spins are in $|\!\!\downarrow\rangle$.

Important examples include the coherent spin state (CSS),
\(
\Big|\frac{N}{2},\frac{N}{2}\Big\rangle
=
|\!\!\uparrow\uparrow\cdots\uparrow\rangle,
\)
and the GHZ state,
\(
|\psi_{{\rm GHZ}_z}\rangle
=
\frac{1}{\sqrt{2}}
\left(
\Big|\frac{N}{2},\frac{N}{2}\Big\rangle
+
\Big|\frac{N}{2},-\frac{N}{2}\Big\rangle
\right).
\)

Nonlinear collective interactions enable the generation of highly nonclassical states within the symmetric spin manifold. In particular, nonlinear dynamics such as OAT, TAT, and TNT can produce strong squeezing and multipartite entanglement, providing versatile resources for both single-parameter and multiparameter quantum metrology \cite{PhysRevResearch.6.033292,RevModPhys.90.035005,10.1063/5.0204102}.
Comprehensive reviews of nonlinear collective-spin dynamics and their applications to quantum-enhanced metrology can be found in Refs.~\cite{RevModPhys.90.035005,10.1063/5.0204102}.

\subsection{GHZ state formation via OAT}
To generate a GHZ state, we apply the OAT evolution along the $x$ axis to an ensemble initialized in the ground state of \( H_0 = \omega J_z\;, \omega > 0 \), namely \( |N/2,-N/2\rangle \). The dynamics are governed by the OAT Hamiltonian \( H_{\text{OAT}}=\chi J_x^2 \), where $\chi$ denotes the nonlinear interaction strength~\cite{PhysRevA.47.5138,MA201189,RevModPhys.90.035005,10.1063/5.0204102}. 
The corresponding time-evolution operator is \( U_{\text{OAT}}(t)=\exp(-i\chi t J_x^2) \).
This dynamics generate quadratic phase accumulation on the collective-spin amplitudes, leading to self-interference in the Dicke basis. 
To see this mathematically, we expand the initial state $|N/2,-N/2\rangle$ in the eigenbasis of $J_x$,
\begin{align}
|\psi_0\rangle = \sum_{m_x} c_{m_x} |m_x\rangle,
\end{align}
where $|m_x\rangle$ denotes the eigenstate of $J_x$ with eigenvalue $m_x$, each eigenstate acquires a quadratic phase under the OAT evolution,
\begin{align}
|\psi(t)\rangle = \sum_{m_x} c_{m_x} e^{-i\chi t m_x^2} |m_x\rangle.
\end{align}
This quadratic phase accumulation generates quantum interference among the Dicke components. At $\chi t = \pi/2$, the phase factor becomes $e^{-i\pi m_x^2/2}$. Since $m_x^2$ is even under $m_x \to -m_x$, symmetric pairs $(m_x, -m_x)$ acquire the same phase. Moreover, $m_x^2 \mod 2$ is periodic, so the phase $e^{-i\pi m_x^2/2}$ takes only two distinct values depending on the parity of $m_x^2$. This produces constructive interference exclusively for the extremal components $m_x = \pm N/2$, while all intermediate components cancel destructively, causing the distribution to collapse into two macroscopically distinct components. 
After a collective rotation $R_y(-\pi/2)$ back to the $z$-basis, the state reduces to
\begin{align}
|\psi(\pi/2\chi)\rangle \propto |\uparrow\uparrow\cdots\uparrow\rangle + e^{i\phi}|\downarrow\downarrow\cdots\downarrow\rangle = |\psi_{{\rm GHZ}_z}\rangle.
\end{align}

As $\chi t$ varies, correlations and interference patterns emerge in phase space. The Husimi $Q$-function provides a convenient visualization of these patterns, while the underlying framework is developed directly in the Dicke basis, where nonlinear dynamics imprint phases on the collective-spin amplitudes. Figure~\ref{fig1}(a) shows the Husimi distributions for different squeezing angles $\chi t$, illustrating the emergence of these interference patterns.

The mathematical underpinning of this mechanism is the Talbot-type fractional revival of quantum states~\cite{AVERBUKH1989449,Gerry1993}, where at $\chi t = \pi/2$ the periodicity of the accumulated phases produces constructive interference at the extremal Dicke components and destructive interference elsewhere~\cite{PhysRevLett.132.113402}.
Phase accumulation driving constructive and destructive interference among collective-spin amplitudes defines the interference-based framework adopted here, consistent with  literature~\cite{PhysRevLett.90.030402,PhysRevB.94.205118,PhysRevLett.132.113402}.

Interestingly, the same OAT dynamics at $\chi t = \pi/2$ is also used in interaction-based readout to map the encoded phase onto a collective population observable~\cite{PhysRevA.98.012129}. While here it generates a GHZ state through interference among Dicke components, in interaction-based readout the same nonlinear phase evolution converts the encoded phase into an experimentally accessible signal. This connection highlights the role of interference in both state preparation and readout.

\subsection{multi-GHZ-like state formation via TAT}
TAT generates squeezing via nonlinear twisting about two orthogonal axes~~\cite{PhysRevA.47.5138,MA201189,10.1063/5.0204102}. Starting from the GHZ state $|\psi_{{\rm GHZ}_z}\rangle$, the system evolves under
\(
H_{\rm TAT}=\chi (J_x^2-J_y^2).
\)
This evolution deforms the two extremal Dicke components in opposite directions, so that the Husimi distribution spreads from the $\pm z$ poles toward the $\pm x$ and $\pm y$ axes~\cite{PhysRevResearch.6.033292}, as shown in Fig.~\ref{fig1}(b).

At the appropriate twisting time, the TAT-driven spreading of the spin distribution leads to overlap and self-interference of the spin components~\cite{PhysRevResearch.6.033292}, and this effect produces a coherent superposition of GHZ components aligned along different axes. The resulting multi-GHZ-like state can be written as
\begin{align}\label{eq:mulQHZ}
|\psi_{\text{mGHZ}}\rangle =
\frac{1}{\mathcal{N}}
\Big(
\big|\psi_{{\rm GHZ}_x}\big\rangle
+
\big|\psi_{{\rm GHZ}_y}\big\rangle
+
\big|\psi_{{\rm GHZ}_z}\big\rangle
\Big),
\end{align}
where $\mathcal{N}$ is a normalization constant and $|\psi_{{\rm GHZ}_\mu}\rangle$ denotes a GHZ state polarized along the $\mu$ axis. This evolution is illustrated in Fig.~\ref{fig1}(b). 

\begin{figure}[t]
    \centering
    \includegraphics[width=\linewidth]{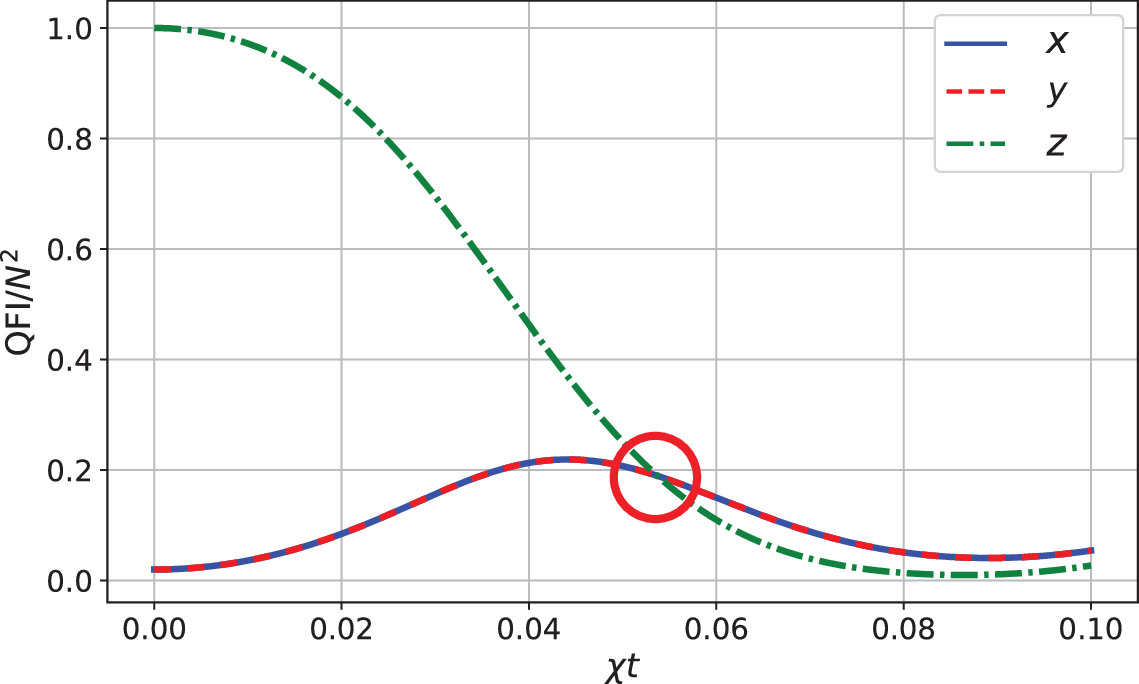}
    \caption{Normalized QFI along each axis $\mu \in \{x, y, z\}$ as a function of the TAT evolution time $\chi t$, for $N = 50$. At the optimal twisting time (circle), $\mathrm{QFI}_x = \mathrm{QFI}_y = \mathrm{QFI}_z$, indicating simultaneous Heisenberg scaling sensitivity along all three axes.}
    \label{fig2}
\end{figure}

Notably, $|\psi_{\text{mGHZ}}\rangle$ has been shown to be an optimal resource for multiparameter estimation~\cite{PhysRevResearch.6.033292}.
To quantify the metrological content of the TAT-generated state, Figure~\ref{fig2} shows the normalized QFI$/N^2$ along each axis $\{x,y,z\}$ as a function of $\chi t$ for $N=50$. At $\chi t = 0$, the initial $|\mathrm{GHZ}_z\rangle$ state concentrates all metrological resource along the $z$ axis (QFI$/N^2 = 1$, Heisenberg limit), while the QFI along $x$ and $y$ remain small, reflecting the trade-off inherent to single-axis probes~\cite{PhysRevA.94.052108}. As the TAT evolution proceeds, the resource is redistributed symmetrically among all three axes. At the optimal time, $\mathrm{QFI}_x = \mathrm{QFI}_y = \mathrm{QFI}_z$ (shown by the red circle), and the QFIM $\mathbf{Q}$ is approximately diagonal with equal entries, so $\mathrm{Tr}[\mathbf{Q}^{-1}]$ is minimized and the state achieves simultaneous Heisenberg-limit precision along all three axes with no trade-off~\cite{PhysRevResearch.6.033292}. The QFI is defined in Sec.~\ref{sec3}, and its enhancement under control is analyzed in Sec.~\ref{sec5}.

\section{Single-parameter metrology with GHZ probes under noise}\label{sec3}
In this section, we analyze the effect of noise on GHZ-based metrology, which provides a reference for the Hamiltonian control strategies developed in the next sections.

\subsection{Metrology scheme for sensing an external field}
We consider a standard quantum metrology protocol where an ensemble of $N$ two-level particles is used to estimate an external field. The probe-field coupling is described by
\(
H(\phi)=\phi J_z,
\)
where $\phi$ is the unknown parameter to be estimated (e.g., a magnetic-field amplitude). During the sensing time, the probe accumulates a $\phi$-dependent phase, which is then inferred from an appropriate measurement.


We initialize the probe in the GHZ state $\rho_0 = |\psi_{{\rm GHZ}_z}\rangle\langle \psi_{{\rm GHZ}_z}|$, which can be prepared via OAT evolution. Here, we assume that the OAT preparation occurs on timescales shorter than the dominant dissipation rates, allowing us to neglect decoherence during state generation and attribute dissipation effects solely to the sensing process.

During the sensing time $t$, the probe state evolves according to the master equation
\begin{align}
\frac{d\rho_\phi(t)}{dt}
=
-i\big[H(\phi),\rho_\phi(t)\big]
+\mathcal{L}\!\left[\rho_\phi(t)\right],
\end{align}
where $\mathcal{L}$ is the Lindblad generator describing decoherence. 
The estimation precision is characterized by the quantum Fisher information (QFI)~\cite{doi:10.1142/S0219749909004839}
\begin{align}
Q(t)
=
2\sum_{k,l}
\frac{\big|\langle k | \partial_\phi \rho_\phi(t) | l \rangle\big|^2}{\lambda_k + \lambda_l},
\end{align}
where $\rho_\phi(t)=\sum_k \lambda_k |k\rangle\langle k|$ is the spectral decomposition and the sum runs over all pairs satisfying $\lambda_k+\lambda_l\neq 0$. The QFI determines the quantum Cram\'er-Rao bound,
\(
\mathrm{Var}(\hat{\phi}) \geq \frac{1}{\nu Q(t)},
\)
where $\nu$ is the number of independent repetitions.

\subsection{The effect of noise}
In practical quantum metrology, the probe is inevitably affected by noise. We consider both local noise and collective noise processes arising from a shared environment, which induce correlated emission, pumping, and dephasing \cite{PhysRevA.98.063815}. These effects degrade quantum coherence and reduce the achievable estimation precision. The probe state evolution explicitly gives
\begin{align}\label{eq:Lindblad_local} \notag\frac{d\rho_\phi(t)}{dt} = -i\Big[H(\phi), \rho_\phi(t)\Big] + &\sum_{n=1}^N\sum_j \gamma_j\mathcal{L}_j[\rho_\phi(t)] \\ 
&+ \sum_j \Gamma_j \mathcal{L}_j[\rho_\phi(t)], \end{align}
where $\gamma_j$ and $\Gamma_j$ denote the dissipation rates of local and collective noise channels, respectively. Each Lindblad term takes the standard form
\(
\mathcal{L}_j[\rho]
=
L_j \rho L_j^\dagger
-\frac{1}{2}\Big\{L_j^\dagger L_j,\rho\Big\},
\)
with $L_j$ is the jump operator. The local and collective noise types considered in this work are summarized in Table~\ref{tab:local}.

\begin{figure}[t]
    \centering
    \includegraphics[width=\linewidth]{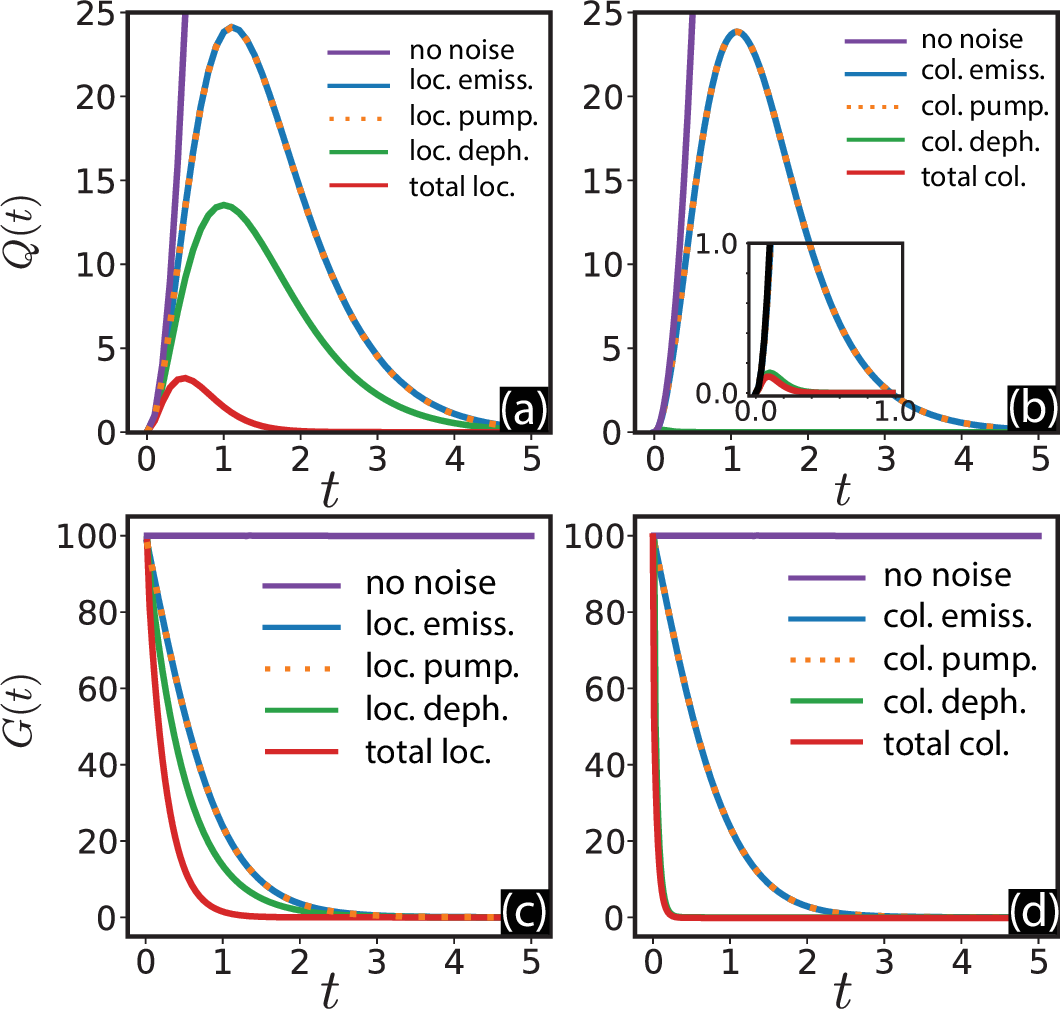}
    \caption{(a,b) The QFI $Q(t)$ as a function of time $t$ under different noise channels. (c,d) The metrological gain $G(t)=Q(t)/t^2$ for the same cases. Parameters: $N=10$, with local dissipation rates $\gamma=0.2$ and collective dissipation rates $\Gamma=0.2$ for each channel.}
    \label{fig3}
\end{figure}

\begin{table}[t!]
\footnotesize
\centering
\caption{Summary of the local and collective noise models \cite{PhysRevA.98.063815}.}
\begin{tabular}{lll}
\hline
\textbf{Noise} & \textbf{Rate} & \textbf{Lindbladian \(\mathcal{L}[\rho]\)} \\
\hline
Local emission & \(\gamma_-\) & \(\sum_n 2\sigma_-^{(n)} \rho \sigma_+^{(n)} - \{ (N/2 - J_z), \rho \}\) \\

Local pumping & \(\gamma_+\) & \(\sum_n 2\sigma_-^{(n)} \rho \sigma_+^{(n)} - \{ (J_z - N/2), \rho \}\) \\

Local dephasing & \(\gamma_z\) & \(\sum_n \frac{1}{2}\sigma_z^{(n)} \rho \sigma_z^{(n)} - \frac{N}{2} \rho\) \\
Collective emission & \(\Gamma_-\) & \(\sum_{m,n} 2\sigma_-^{(m)} \rho \sigma_+^{(n)} - J_+ J_- \rho - \rho J_+ J_-\)\\
Collective pumping & \(\Gamma_+\) & \(\sum_{m,n} 2\sigma_-^{(m)} \rho \sigma_+^{(n)} - J_- J_+ \rho - \rho J_+ J_-\) \\
Collective dephasing & \(\Gamma_z\) & \(\sum_{m,n} \frac{1}{2}\sigma_z^{(m)} \rho \sigma_z^{(n)} - J_z^2 \rho - \rho J_z^2\)\\
\hline
\end{tabular}
\label{tab:local}
\end{table}

\begin{figure*}[t]
    \centering
    \includegraphics[width=0.9\textwidth]{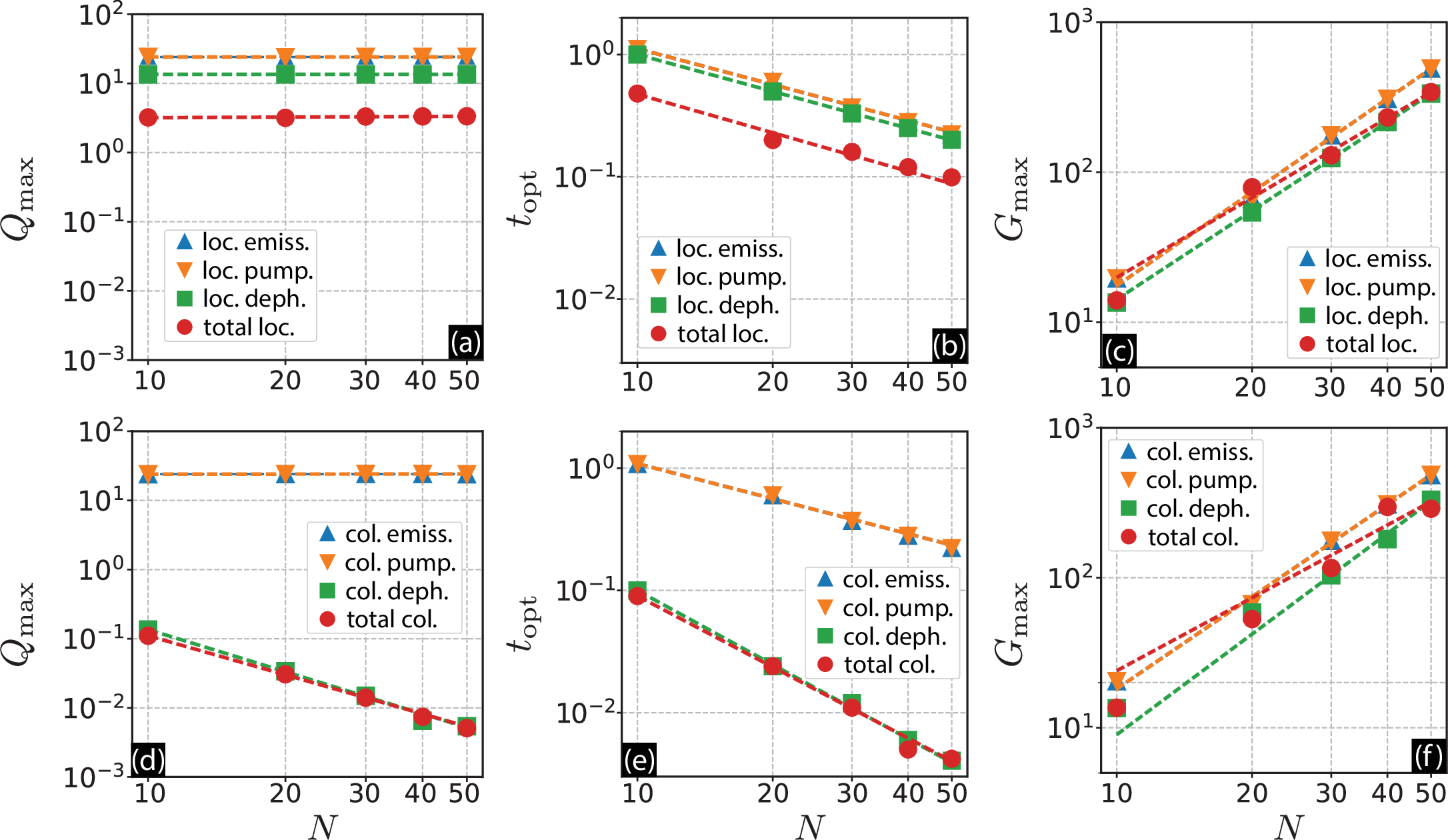}
    \caption{(a) Maximum QFI \( Q_{\text{max}} \) under local noise as a function of the spin number \( N \). (b) Optimal sensing time \( t_{\text{opt}} \) corresponding to \( Q_{\text{max}} \). (c) Maximum metrological gain \( G_{\text{max}} = Q_{\text{max}} / t_{\text{opt}}^2 \). (d-f) Same quantities as in panels (a-c), but for collective noise.
    }
    \label{fig4}
\end{figure*}

We analyze the QFI under different noise channels in Fig.~\ref{fig3}(a,b). In the noiseless case, the QFI scales as $Q(t)=N^2t^2$, showing ideal Heisenberg limit. In contrast, under noise, the QFI increases only at short times, reaches a maximum at an optimal sensing time, and then decreases as decoherence suppresses coherence and correlations.

We next evaluate the metrological gain $G(t)=Q(t)/t^2$, shown in Fig.~\ref{fig3}(c,d). In the noiseless case, $G(t)$ remains constant at the Heisenberg value $G(t)=N^2$ (here $N=10$). Under noise, $G(t)$ decreases with time and then vanishes, indicating the loss of quantum-enhanced sensitivity.

Moreover, emission and pumping show similar behavior for both local and collective noise, since they are symmetric population-transfer processes. In contrast, dephasing has a distinct impact: it causes a much faster decay of $G(t)$ because it directly destroys the phase coherence required for estimation. See App.~\ref{app:spin_dynamics} for further details.

\subsection{The effect of spin number $N$}
Figure~\ref{fig4} summarizes the system-size scaling of the maximum QFI ($Q_{\rm max}$), the optimal sensing time ($t_{\rm opt}$), and the maximum metrological gain ($G_{\rm max}=Q_{\rm max}/t_{\rm opt}^2$) under different noise channels. Panels (a-c) show the local-noise case, while panels (d-f) show the collective-noise case.

For local noise, $Q_{\rm max}$ [Fig.~\ref{fig4}(a)] is independent of $N$. This is because the local decoherence on each spin rapidly suppresses the GHZ coherence term $|\!\!\uparrow\cdots\uparrow\rangle\langle\downarrow\cdots\downarrow\!\!|$, which is responsible for the ideal Heisenberg scaling. As a result, $Q_{\rm max}$ does not increase with system size (see also Fig.~\ref{fig10} in App.~\ref{app:spin_dynamics}). Meanwhile, $t_{\rm opt}$ [Fig.~\ref{fig4}(b)] decreases with increasing $N$ since larger ensembles lose coherence more rapidly. Consequently, $G_{\rm max}$ follows an approximate power-law scaling $G_{\rm max}\propto N^\alpha$, mainly driven by the reduction of the optimal sensing time.

For collective noise [Figs.~\ref{fig4}(d--f)], the scaling depends strongly on the channel. Under the collective emission and pumping, $Q_{\rm max}$ remains large and independent of $N$, and is almost identical to the local-noise case at the same dissipation rate, indicating that noise correlations have little effect in this regime. In contrast, collective dephasing strongly suppresses $Q_{\rm max}$ because random phase rotations generated by $J_z$ rapidly destroy the GHZ coherence. In all collective-noise cases, $t_{\rm opt}$ decreases faster with $N$ than for local noise, leading to an overall increase of $G_{\rm max}$ with system size. These results show that noise correlations can significantly modify the system-size scaling of metrological performance.

\begin{figure*}[t]
    \centering
    \includegraphics[width=0.9\linewidth]{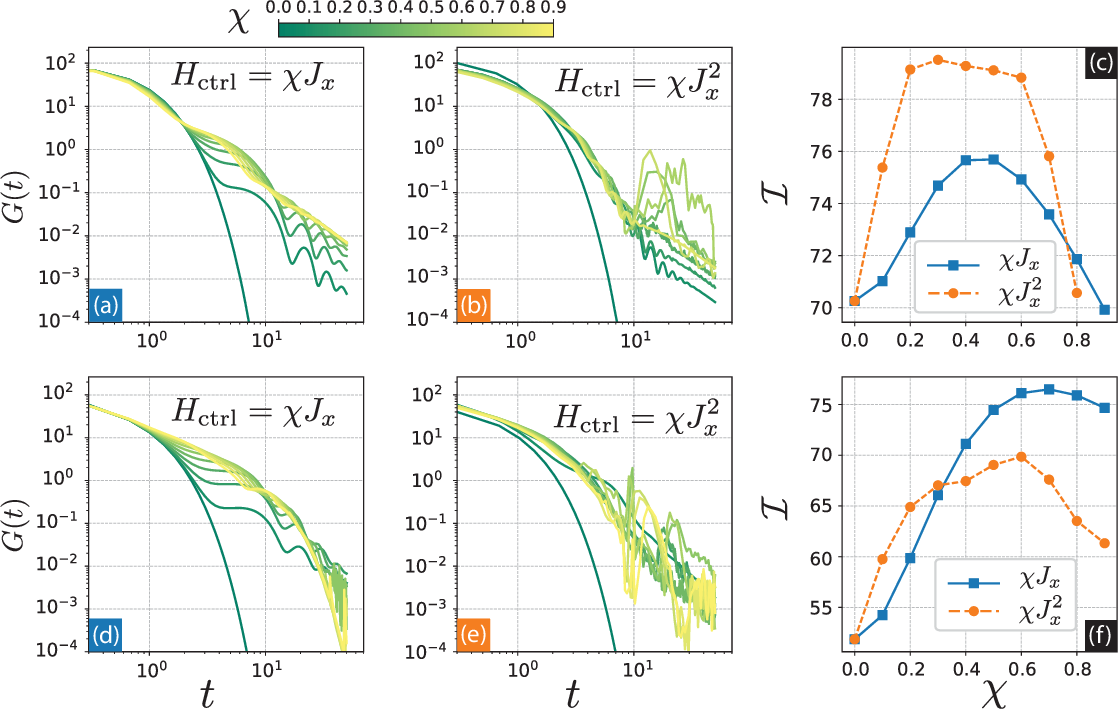}
    \caption{Metrological gain under the local noise. Panels (a-c) show the results for the local emission, while panels (d-f) correspond to the local dephasing. In (a,b) and (d,e), the metrological gain \(G(t)\) is plotted for different control strengths \(\chi\) (color-coded), comparing the linear control Hamiltonian \(H_{\rm ctrl}=\chi J_x\) and the nonlinear control \(H_{\rm ctrl}=\chi J_x^2\). (c,f) show the integrated gain \(\mathcal{I}=\int_0^T G(t)\) as a function of \(\chi\), for a truncation of $T = 50$.
}
    \label{fig5}
\end{figure*}

\section{Control-assisted metrology via Hamiltonian engineering}\label{sec4}
Noise reduces the coherence of the probe state, which directly lowers the QFI and degrades the estimation precision. To mitigate this decoherence and enhance metrological sensitivity, quantum control can be applied during the sensing dynamics to stabilize the probe evolution and enhance the useful quantum correlations.

To mitigate noise-induced decoherence, we introduce a control Hamiltonian during the sensing dynamics. The total Hamiltonian is written as
\begin{align}
H_{\text{total}}(\phi)=H(\phi)+H_{\text{ctrl}},
\end{align}
where $H_{\text{ctrl}}$ is an externally applied control term used to protect coherence and enhance the QFI.

Here, we consider several control Hamiltonians relevant for noisy spin sensing, including a linear interaction $H_{\text{ctrl}}=\chi J_x$ and a nonlinear interaction $H_{\text{ctrl}}=\chi J_x^2$, where $\chi$ is the control strength.
Both are standard tools in collective-spin platforms. The linear drive $\chi J_x$ is realized as a global transverse field in atomic ensembles and optical clocks~\cite{Pedrozo2020,Robinson2024}. The nonlinear interaction $\chi J_x^2$ has been demonstrated in cavity-QED setups, Bose-Einstein condensates, and atom-chip experiments~\cite{PhysRevLett.104.073602,PhysRevA.81.021804,Riedel2010,Chalopin2018}, though it requires an effective all-to-all spin-spin coupling, which is more resource-intensive than a simple driving field and can be engineered via cavity-mediated interactions or collisional interactions in BECs at the cost of additional experimental overhead. Both controls are physically motivated as $J_x$ partially averages out phase noise via global rotation, while $J_x^2$ builds up useful correlations before decoherence dominates.
Including these controls, the probe state is governed by the Lindblad master equation in Eq.~\eqref{eq:Lindblad_local}, with $H(\phi)$ replaced by $H_{\rm total}(\phi)$. The control strength $\chi$ is scanned numerically to identify optimal operating points, which can partially protect coherence and generate noise-resilient correlations, leading to an enhanced QFI. The entire protocol is open-loop, with no adaptive feedback involved.

\subsection{For local noise}
Figure~\ref{fig5} summarizes the metrological performance of the controlled protocol in the presence of local noise. We quantify the sensing advantage using the metrological gain $G(t)$ and its time-integrated value
\(
\mathcal{I}=\int_{0}^{\infty}G(t)dt,
\)
which captures the overall enhancement accumulated over the sensing window $[0,\infty)$.
In numerical simulations, the integration is truncated at \(T=50\), which is sufficient to capture the decay of all curves down below \(10^{-2}\).

We first consider local emission (with similar behavior for pumping), shown in Figs.~\ref{fig5}(a-c). Panels (a) and (b) plot the gain $G(t)$ for different control strengths $\chi$, comparing the linear drive $H_{\rm ctrl}=\chi J_x$ in (a) with the nonlinear control $H_{\rm ctrl}=\chi J_x^2$ in (b). In both cases, applying control slows down the decay of $G(t)$ compared to the uncontrolled dynamics, indicating partial protection against emission-induced decoherence. Increasing $\chi$ enhances the short-time gain. The integrated gain $\mathcal{I}$ is shown in panel (c), where an optimal regime appears at intermediate $\chi$. Notably, the nonlinear control yields a significantly larger $\mathcal{I}$ than the linear drive.

\begin{figure*}[t]
    \centering
    \includegraphics[width=0.9\linewidth]{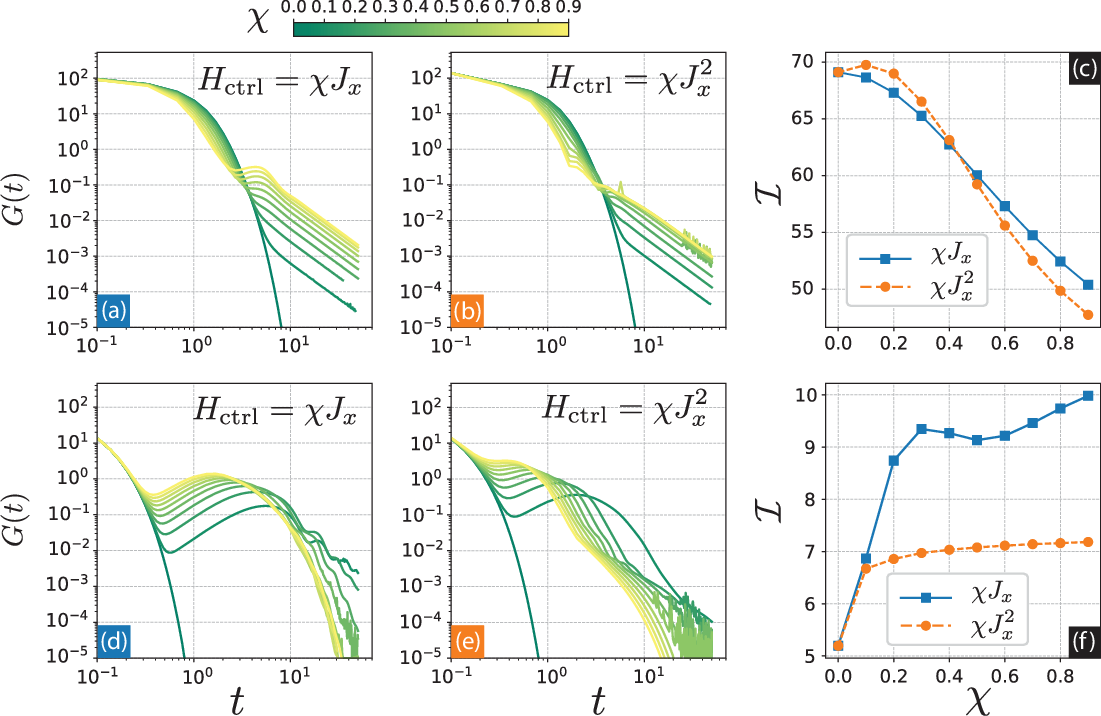}
    \caption{Metrological gain under the collective noise. (a-c) show the results for the collective emission, while (d-f) correspond to the collective dephasing. Panels (a,b) and (d,e) display the metrological gain \(G(t)\) for different control strengths \(\chi\) (color-coded), comparing the linear control Hamiltonian \(H_{\rm ctrl}=\chi J_x\) and the nonlinear control Hamiltonian \(H_{\rm ctrl}=\chi J_x^2\). (c,f) show the integrated gain $\mathcal{I}$ as a function of \(\chi\). 
}
    \label{fig6}
\end{figure*}

We next consider the local dephasing, shown in Figs.~\ref{fig5}(d-f). Panels (d) and (e) plot $G(t)$ for $H_{\rm ctrl}=\chi J_x$ and $H_{\rm ctrl}=\chi J_x^2$, respectively. Compared with the emission case, the dephasing suppresses the gain more strongly and reduces the benefit of the nonlinear control. As shown in panel (f), the integrated gain $\mathcal{I}$ increases monotonically with $\chi$ for the linear control, whereas the nonlinear control yields only a limited improvement. This indicates that the optimal control depends on the dominant noise channel: the nonlinear control is advantageous under the emission, while the linear drive becomes more effective under the dephasing.

These behaviors can be understood from the interplay between the control Hamiltonian and the dominant decoherence mechanism. 
Under the local emission (and similarly pumping), the decay is mainly due to the population relaxation. Moderate control can accelerate the buildup of useful correlations at short times, leading to an optimal $\chi$ in the integrated gain $\mathcal{I}$, while overly strong control becomes ineffective. Under the local dephasing, the coherence is directly suppressed, and the linear drive $H_{\rm ctrl}=\chi J_x$ is more effective because it continuously rotates the collective spin and partially averages out phase noise, whereas the nonlinear interaction provides only limited improvement.

\subsection{For collective noise}
Figure~\ref{fig6} summarizes the effect of quantum control on the metrological gain under the collective noise. We first consider the collective emission, shown in Figs.~\ref{fig6}(a-c). Panels (a) and (b) plot $G(t)$ for different control strengths $\chi$, comparing $H_{\rm ctrl}=\chi J_x$ with $H_{\rm ctrl}=\chi J_x^2$. In both cases, $G(t)$ decays rapidly due to collective relaxation. Increasing $\chi$ slightly modifies the short- and intermediate-time behavior, but reduces the overall performance. This is reflected in Fig.~\ref{fig6}(c), where $\mathcal{I}$ decreases monotonically with $\chi$, indicating that strong control is not beneficial under the collective emission.

We next examine the collective dephasing, shown in Figs.~\ref{fig6}(d-f). Here, $G(t)$ exhibits a pronounced plateau at intermediate times, approximately in the range \(t\sim 0.5-5\), whose magnitude depends on $\chi$. In particular, the linear control significantly enhances $G(t)$ over a broad time interval, leading to a clear increase in $\mathcal{I}$. As shown in Fig.~\ref{fig6}(f), $\mathcal{I}$ grows with $\chi$ and saturates for $\chi\gtrsim 0.3$. By contrast, the nonlinear control provides only a modest improvement and saturates at a smaller value.

These results demonstrate that the optimal control strategy depends crucially on the dominant collective noise mechanism. Under the collective emission, applying control does not improve the metrological performance and can even reduce it at large $\chi$, whereas under the collective dephasing, the linear control provides a robust enhancement and yields the largest integrated gain.

These results can be explained as follows. Under the collective emission, the relaxation is governed by collective jump operators, and strong control competes with this correlated dissipation, reducing the overall gain and causing $\mathcal{I}$ to decrease with $\chi$.
Under the collective dephasing, the dominant effect is the correlated phase noise generated by $J_z$. Here, the linear control efficiently counteracts the dephasing-induced loss of coherence and produces a broad enhancement of $G(t)$, while the nonlinear control yields only a modest improvement.

\section{Multiparameter sensing with multi-GHZ-like probes}\label{sec5}
We consider a multiparameter quantum metrology protocol where an ensemble of $N$ spin-$\frac{1}{2}$ particles is used to estimate a three-dimensional magnetic field $\boldsymbol{\phi}=(\phi_x,\phi_y,\phi_z)$. The field couples to the collective spin via
\begin{align}
H(\boldsymbol{\phi}) = 
\boldsymbol{\phi}\cdot\mathbf{J} =
\phi_x J_x+\phi_y J_y+\phi_z J_z,
\end{align}
where $\mathbf{J}=(J_x,J_y,J_z)$ denotes the total spin operator. During the sensing time $t$, the probe accumulates field-dependent phases that encode the three components of $\boldsymbol{\phi}$.

Starting from the initial multi-GHZ-like state $\rho_0$ in Eq.~\eqref{eq:mulQHZ}, the system evolution in the presence of decoherence is governed by the Lindblad master equation
\begin{align}
\frac{d\rho_{\boldsymbol{\phi}}(t)}{dt}
=-i\big[H(\boldsymbol{\phi}),\rho_{\boldsymbol{\phi}}(t)\big]
+\mathcal{L}\left[\rho_{\boldsymbol{\phi}}(t)\right],
\end{align}
where $\mathcal{L}$ denotes the Lindblad generator.

The achievable precision is characterized by the quantum Fisher information matrix (QFIM) $\mathbf{Q}$. Using the spectral decomposition, i.e.,
\(\rho_{\boldsymbol{\phi}}(t)=\sum_{n}\lambda_n\,|\lambda_n\rangle\langle \lambda_n|,
\)
the QFIM elements are given by
\begin{align}
\big[\mathbf{Q}\big]_{\mu\nu}
=
2\sum_{n,m}
\frac{\langle \lambda_n|\partial_{\phi_\mu}\rho_{\boldsymbol{\phi}}(t)|\lambda_m\rangle
\langle \lambda_m|\partial_{\phi_\nu}\rho_{\boldsymbol{\phi}}(t)|\lambda_n\rangle}
{\lambda_n+\lambda_m},
\end{align}
where the sum runs over all eigenpairs with $\lambda_n+\lambda_m>0$ and $\mu,\nu\in\{x,y,z\}$.

For any unbiased estimator $\hat{\boldsymbol{\phi}}$, the covariance matrix satisfies the multiparameter quantum Cram\'er--Rao bound (QCRB),
\begin{align}
\mathrm{Cov}(\hat{\boldsymbol{\phi}})
\succeq
\frac{1}{M}\mathbf{Q}^{-1},
\end{align}
where $M$ is the number of independent repetitions. To quantify the overall estimation performance, we use the weighted cost
\(
\mathcal{C}
=
\mathrm{Tr}\!\left(\mathbf{W}\,\mathrm{Cov}(\hat{\boldsymbol{\phi}})\right),
\)
which obeys the weighted QCRB
\begin{align}
\mathcal{C}
\ge
\frac{1}{M}\mathrm{Tr}\!\left(\mathbf{W}\,\mathbf{Q}^{-1}\right),
\end{align}
with $\mathbf{W}\succeq 0$ a weight matrix. In general, this bound may not be saturable due to measurement incompatibility among different parameters. A sufficient condition for asymptotic compatibility is
\(
\mathrm{Tr}\!\left(\rho_{\boldsymbol{\phi}}(t)\,[L_\mu,L_\nu]\right)=0,
\)
where the symmetric logarithmic derivatives $L_\mu$ are defined by
\(
\partial_{\phi_\mu}\rho_{\boldsymbol{\phi}}(t)
=
\frac{1}{2}\big(L_\mu\rho_{\boldsymbol{\phi}}(t)+\rho_{\boldsymbol{\phi}}(t)L_\mu\big).
\)
In the following, we study the impact of noise on the weighted QCRB.

\subsection{Estimation of the weighted QCRB under different noise channels}

\begin{figure}[t]
    \centering
    \includegraphics[width=\linewidth]{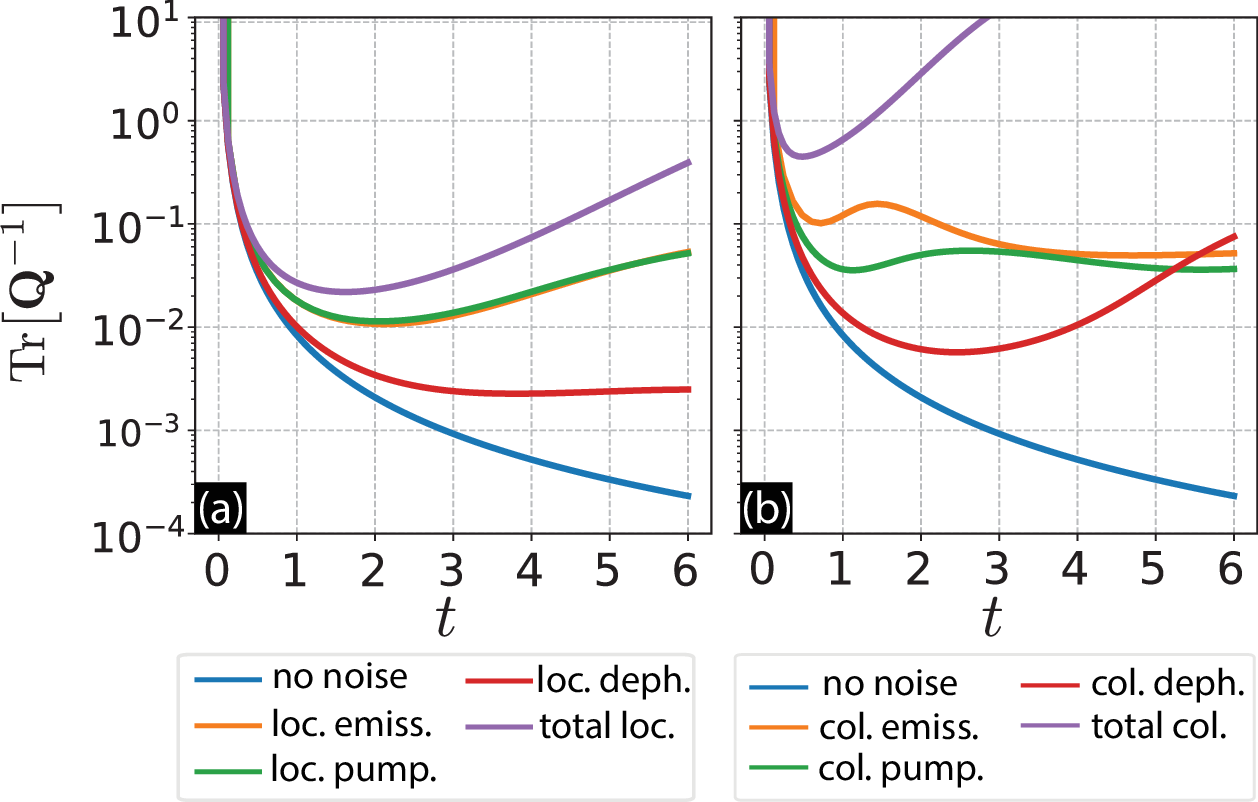}
    \caption{Weighted quantum Cram{\'e}r-Rao bound (weighted QCRB) as a function of interaction time $t$ under (a) local and (b) collective noise channels. Results are shown for emission, pumping, dephasing, and the combined noise case, together with the noiseless reference. In all noisy cases, the bound exhibits a minimum at an optimal time.}
    \label{fig7}
\end{figure}

In our numerics, we set equal weight $\mathbf{W}=\mathbf{I}$ and consider a single repetition ($M=1$), so that the weighted QCRB reduces to $\mathrm{Tr}[\mathbf{Q}^{-1}]$.
Figure~\ref{fig7} shows the weighted QCRB as a function of time $t$ for different noise channels,  as considered in Sec.~\ref{sec3}. In the numerical simulations, the noise strength is fixed at $0.1$ for $N=10$ qubits with and sensing phases $\phi_x=\phi_y=\phi_z=\pi/6$. Figure~\ref{fig7}(a) corresponds to local noise (local emission, pumping, and dephasing), while Fig.~\ref{fig7}(b) shows the corresponding collective noise models. The noiseless case is included for reference.

In the noiseless case, the bound decreases monotonically with $t$. In contrast, under noise it exhibits a clear minimum at an optimal interaction time due to the competition between phase accumulation and noise-induced loss of coherence \cite{Ho2023}.

For the local noise [Fig.~\ref{fig7}(a)], the local dephasing yields the smallest bound, whereas the local emission and pumping show similar but larger values. When all local channels are present, the bound increases significantly and long times become ineffective. For the collective noise [Fig.~\ref{fig7}(b)], the collective emission and pumping behave similarly, while the collective dephasing strongly degrades the precision at long times. The combined collective noise case yields the largest bound, indicating the strongest limitation from simultaneous dissipation.

\subsection{Quantum control on the weighted QCRB under noise}
In the previous sections, the control Hamiltonians were chosen based on
physical intuition specific to each noise channel: linear interaction $\chi J_x$
and a nonlinear interaction $\chi J_x^2$. While these single-axis controls
are physically transparent, they are not guaranteed to be optimal, 
particularly for multiparameter sensing, where the three phases
$(\phi_x, \phi_y, \phi_z)$ are estimated simultaneously and the optimal
control direction may not align with any single axis. Moreover, the
single-axis approach requires separate physical reasoning for each noise
type, which does not generalize easily.

Here we take a general approach, where we optimize over the parameter space of linear and quadratic controls,
\begin{align}
    H_{\rm ctrl}^{\rm lin}  &= \chi_x J_x + \chi_y J_y + \chi_z J_z = \bm\chi\cdot\bm J,
    \label{eq:ctrl_lin} \\
    H_{\rm ctrl}^{\rm quad} &= \chi_x J_x^2 + \chi_y J_y^2 + \chi_z J_z^2 = \bm\chi\cdot\bm J^2.
    \label{eq:ctrl_quad}
\end{align}
To keep the search tractable, we adopt a two-step strategy:
(i)~identify the optimal sensing time $t_{\rm opt}$ from the uncontrolled noisy
curve, then (ii)~optimize $(\chi_x,\chi_y,\chi_z)$ only at $t_{\rm opt}$, reducing
the search to a single-point cost function. The resulting optimal control is then applied over the full time range.

The three parameters $(\chi_x, \chi_y, \chi_z)$ are optimized as follows. A coarse $5^3$ grid scan over the parameter space is first performed to identify the region of best performance, avoiding local minima that would trap a gradient-based optimizer. Starting from the best grid point, the Nelder-Mead method is then applied to refine the solution without requiring gradient information, which is advantageous since $\mathrm{Tr}[\mathbf{Q}^{-1}]$ has no simple analytic form. 

The linear control $H_{\rm ctrl}^{\rm lin}$  ~\eqref{eq:ctrl_lin} is equivalent to a continuous
rotation at rate $\Omega=\|(\chi_x,\chi_y,\chi_z)\|$ around a tunable
axis. For local dephasing, when $\Omega\gg\gamma$ this realizes continuous
dynamical decoupling (CDD), suppressing the effective dephasing rate to
$\gamma_{\rm eff}\approx\gamma^2/(4\Omega^2)\ll\gamma$~\cite{PhysRevLett.82.2417}.
The quadratic control $H_{\rm ctrl}^{\rm quad}$ ~\eqref{eq:ctrl_quad} generates spin squeezing within
the symmetric subspace~\cite{PhysRevA.47.5138}. 

\begin{figure}[t]
    \centering
    \includegraphics[width=\linewidth]{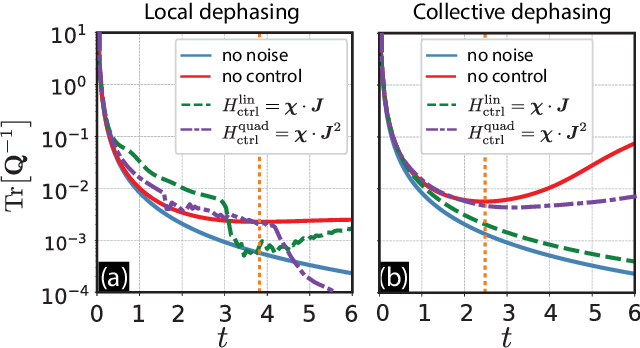}
    \caption{Weighted QCRB $\mathrm{Tr}[\mathbf{Q}^{-1}]$ vs.\ sensing
    time $t$ for $N=10$, $\gamma=0.1$.
    (a)~Local dephasing.
    (b)~Collective dephasing.
    In each panel, blue solid: no noise; red solid: dephasing noise, no control;
    green dashed: linear control $H_{\rm ctrl}^{\rm lin} =\boldsymbol{\chi}\cdot\mathbf{J}$;
    purple dash-dot: quadratic control
    $H_{\rm ctrl}^{\rm quad} =\boldsymbol{\chi}\cdot\mathbf{J}^2$.
    The orange dotted vertical line marks the optimal sensing time $t_{\rm opt}$
    at which the control parameters were optimized.}
    \label{fig8}
\end{figure}

Figure~\ref{fig8} shows numerical results of $\mathrm{Tr}[\mathbf{Q}^{-1}]$. For local dephasing (a), the result from $H_{\rm ctrl}^{\rm quad}$ closely follows the no-control curve and gradually reduces the variance below it after $t_{\rm opt}$, while the result from $H_{\rm ctrl}^{\rm lin}$ remains above the no-control curve for most of the evolution, exhibiting a dip near $t_{\rm opt}$ before rising again, indicating that spin squeezing provides a more stable long-time improvement. For collective dephasing (b), $H_{\rm ctrl}^{\rm lin}$ dramatically reduces the variance well all the time, while $H_{\rm ctrl}^{\rm quad}$ provides only moderate improvement, as the linear control effectively decouples the correlated phase noise via global rotation, whereas the quadratic control is limited by $[J_z^2, J_z]=0$.

\section{Conclusion}\label{sec6}
We have developed an interference-based description of collective spin-$\frac{1}{2}$ dynamics and used it to elucidate the formation of metrologically useful entangled states under nonlinear twisting interactions. Within this framework, the OAT and TAT naturally generate GHZ and multi-component GHZ structures through nonlinear phase accumulation and self-interference inside the Dicke manifold, providing a transparent physical interpretation of state formation.

We then examined how local and collective noise processes modify these interference patterns and impose a finite optimal sensing time. By evaluating the quantum Fisher information and the corresponding metrological gain, we showed that decoherence progressively suppresses the interference features responsible for enhanced sensitivity. In the single-parameter setting, Hamiltonian-level control can partially mitigate this degradation, with performance strongly dependent on the dominant noise channel. Nonlinear control proves more effective in the presence of emission-type noise, whereas linear driving offers greater robustness under dephasing. For three-parameter magnetic-field estimation, the optimized Hamiltonian control yields quantum enhancement, demonstrating that multiparameter sensing benefits from control strategies that respect the symmetry of the estimation problem.
Our results establish a unified connection between nonlinear interference, decoherence, and Hamiltonian control in collective-spin metrology, and provide quantitative benchmarks for realistic sensing platforms.

Future work may explore extensions to non-Markovian noise regimes and more complex sensing scenarios beyond collective spin ensembles.

\acknowledgements
We acknowledge financial support from KAKENHI projects No. JP24K00563, No. JP24K02948, and No. JP24K06878. HM also acknowledges support from CSIS, Tohoku University. LBH is supported by the Tohoku Initiative for Fostering Global Researchers for Interdisciplinary Sciences (TI-FRIS) of MEXT's Strategic Professional Development Program for Young Researchers.

\section*{Data availability}
The data that support the findings of this article are openly
available \cite{Ho2026}.

\section*{Conflict of interest}
The authors declare that there are no conflicts of interest associated with this study.

\begin{widetext}
\appendix

\renewcommand{\theequation}{A.\arabic{equation}}
\setcounter{equation}{0}

\section{Dynamic of spin system under noise.}\label{app:spin_dynamics}
We consider a single spin-$\tfrac{1}{2}$ particle coupled to a noisy environment.
The system’s density matrix \(\rho(t)\) evolves according to the Lindblad master equation:
\begin{equation}
\dot{\rho}(t) = -i[H, \rho(t)] + \sum_{k} \mathcal{D}[L_k]\rho(t),
\label{eq:A.1}
\end{equation}
where \(H\) is the system Hamiltonian, \(L_k\) are the Lindblad operators describing various decoherence channels, and the dissipator is defined as
\begin{equation}
\mathcal{D}[L_k]\rho = L_k \rho L_k^\dagger - \tfrac{1}{2}{L_k^\dagger L_k, \rho}.
\label{eq:A.2}
\end{equation}
We analyze below the characteristic dynamics under several physically relevant types of noise: amplitude damping, excitation (pumping), and dephasing.

For a single spin in an external field, the Hamiltonian is typically
\begin{equation}
H(\phi) = \phi\sigma_z.
\label{eq:A.3}
\end{equation}
In the absence of noise, the Bloch vector \(\vec{r}(t)\) rotates uniformly around the \(z\)-axis:
\begin{align}
\rho(t) &= e^{-iHt}\rho(0)e^{iHt}, \\
\vec{r}(t) &= (r_x\cos\omega_0 t - r_y\sin\omega_0 t,, r_x\sin\omega_0 t + r_y\cos\omega_0 t,, r_z).
\end{align}
Thus, the population remains constant while the phase coherence oscillates periodically.

\subsection{Emission noise}
Emission corresponds to spontaneous relaxation from the excited state \(|1\rangle\) to the ground state \(|0\rangle\), with Lindblad operator
\begin{equation}
L = \sqrt{\gamma_-}\sigma_-
\label{eq:A.4}
\end{equation}
where \(\sigma_- = |0\rangle\langle1|\) and \(\gamma_-\) is the decay rate.

The master equation becomes
\begin{align}
    \dot{\rho} = -i[H(\phi),\rho] + \gamma_- \Big( \sigma_- \rho \sigma_+ - \tfrac{1}{2}\{\sigma_+\sigma_-, \rho\} \Big).
\end{align}
In Bloch-vector form \(\rho = \tfrac{1}{2}(I + \vec{r}\cdot\vec{\sigma})\), the dynamics are
\begin{align}
\dot{r}_x = -\frac{\gamma_-}{2} r_x - \phi r_y, \qquad
\dot{r}_y = \phi r_x - \frac{\gamma_-}{2} r_y, \qquad
\dot{r}_z = -\gamma_- (r_z + 1).
\end{align}
The steady state is \(\vec{r}_{\text{ss}} = (0, 0, -1)\), corresponding to the ground state \(|0\rangle\).
Amplitude damping thus reduces excited-state population and coherence simultaneously, leading to irreversible energy loss. 
See Fig.~\ref{fig9} for the state evolution.

\begin{figure}[b]
    \centering
    \includegraphics[width=0.6\linewidth]{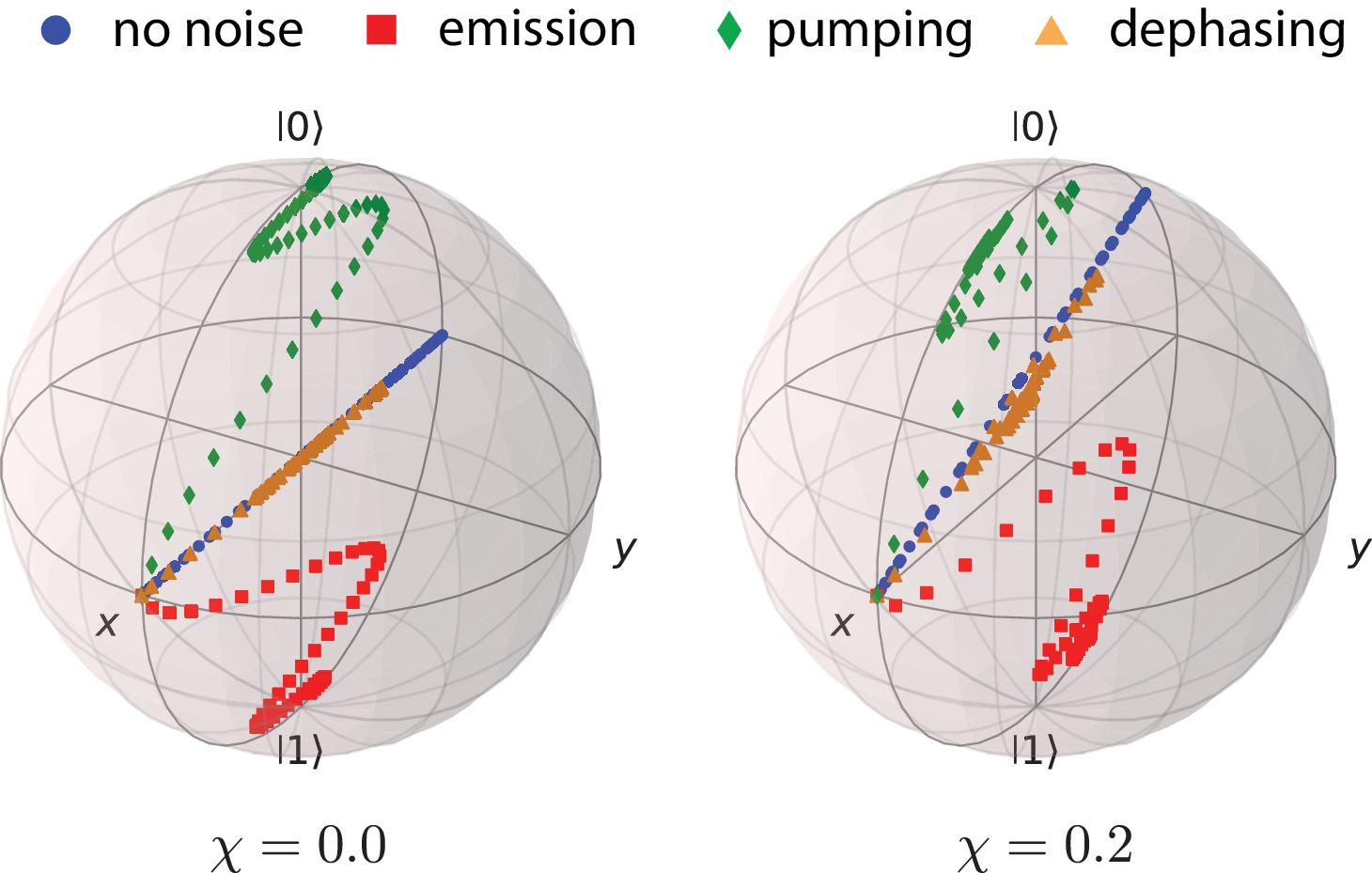}
    \caption{Bloch-sphere trajectories illustrating the time evolution of a single spin subject to various noise types, comparing the cases without control \((\chi = 0)\) and with control \((\chi = 0.2)\).}
    \label{fig9}
\end{figure}

\subsection{Pumping}
The inverse process, thermal pumping or excitation, drives the system toward \(|1\rangle\).
The corresponding Lindblad operator is
\begin{equation}
L = \sqrt{\gamma_+}\sigma_+,
\label{eq:A.5}
\end{equation}
with \(\sigma_+ = |1\rangle\langle0|\).
The Bloch dynamics are
\begin{align}
\dot{r}_x = -\frac{\gamma_+}{2}r_x - \phi r_y, \qquad
\dot{r}_y = \phi r_x - \frac{\gamma_+}{2}r_y, \qquad
\dot{r}_z = \gamma_+(1 - r_z),
\end{align}
yielding the steady state \(\vec{r}_{\text{ss}} = (0,0,+1)\).
This process can model incoherent excitation due to an external pump field or a thermal bath at high temperature.
See Fig.~\ref{fig9} for the state evolution.

\subsection{Dephasing}
Dephasing destroys coherence without changing population.
The corresponding Lindblad operator is
\begin{equation}
L = \sqrt{\gamma_\phi}\sigma_z,
\label{eq:A.6}
\end{equation}
where \(\gamma_\phi\) is the dephasing rate.
The evolution equations are
\begin{align}
\dot{r}_x = -2\gamma_\phi r_x - \phi r_y, \qquad
\dot{r}_y = \phi r_x - 2\gamma_\phi r_y, \qquad
\dot{r}_z = 0.
\end{align}
The steady state retains its population but loses phase information \((r_x, r_y \to 0)\).
In other words, the density matrix becomes diagonal in the energy basis, representing a classical mixture of \(|0\rangle\) and \(|1\rangle\).
See Fig.~\ref{fig9} for the state evolution.

\subsection{Combined effects}
In realistic systems, all three processes may coexist
\begin{equation}
\dot{\rho} = -i[H,\rho] + \gamma_- \mathcal{D}[\sigma_-]\rho + \gamma_+ \mathcal{D}[\sigma_+]\rho + \gamma_\phi \mathcal{D}[\sigma_z]\rho.
\label{eq:A.7}
\end{equation}
This equation describes the most general Markovian noise affecting a single spin.
The steady state corresponds to a thermal equilibrium distribution,
\begin{align}
    \rho_{\text{ss}} = \frac{1}{2}\big(I + r_z^{\text{(ss)}} \sigma_z\big), \quad
r_z^{\text{(ss)}} = \frac{\gamma_+ - \gamma_-}{\gamma_+ + \gamma_-}.
\end{align}
The emission noise dominates at low temperature \((\gamma_+ \ll \gamma_-)\), while pumping dominates at high temperature \((\gamma_+ \gg \gamma_-)\).
Dephasing accelerates the decay of off-diagonal elements, reducing quantum coherence and thus suppressing the QFI and metrological advantage.

\subsection{Block sphere presentation}
Figure~\ref{fig9} show the Bloch-sphere trajectories of the evolution of a single-spin state under different noise channels. Each curve corresponds to a distinct noise process: emission (red squares), pumping (green diamonds), dephasing (orange triangles), and the noiseless case (blue circles).

The left panel illustrates the dynamics without control \((\chi = 0)\), where noise drives the state away from the ideal trajectory and gradually destroys coherence. Emission and pumping processes lead to relaxation toward \(|1\rangle\) and \(|0\rangle\), respectively, while dephasing causes the trajectory to collapse toward the (z)-axis, erasing the azimuthal phase information.

In contrast, the right panel presents the controlled case with a linear control Hamiltonian \(H_{\rm ctrl} = \chi \sigma_x\) for \(\chi = 0.2\). The applied control field counteracts part of the noise-induced decoherence, maintaining the trajectories closer to the noiseless evolution and preserving larger Bloch-vector amplitudes. This stabilization effect slows down the relaxation toward the steady state, thereby prolonging quantum coherence. As a result, the QFI and the corresponding metrological gain are enhanced. We also note that in this single-spin example, the collective and local noise channels are equivalent since only one spin is considered.

\begin{figure*}[t]
    \centering
    \includegraphics[width=\linewidth]{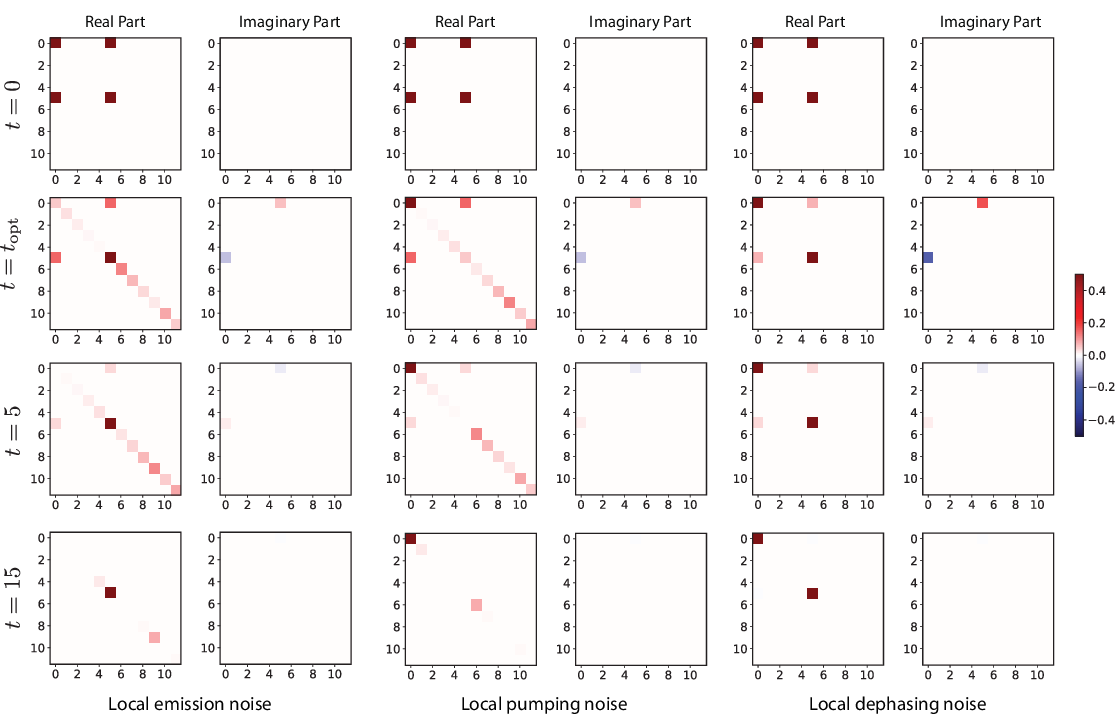}
    \caption{Time evolution of the probe-state density matrix elements for \(N = 5\), \(\gamma = 0.2\). The initial GHZ state \(\rho_{{\rm GHZ}_z}\) occupies only the corner elements \(\rho_{0,0}, \rho_{0,N}, \rho_{N,0},\) and \(\rho_{N,N}\). As the evolution proceeds, population spreads to other elements and Dicke blocks depending on the noise channel. In the long-time limit, the state converges to \(\rho_{N,N}\) (local emission), \(\rho_{0,0}\) (local pumping), or \(\rho_{0,0}+\rho_{N,N}\) (local dephasing).}
    \label{fig10}
\end{figure*}

\subsection{Matrix state presentation}
To illustrate the dynamics for a larger spin system, we show the time evolution of the density matrix elements for \(N = 5\) spins with a noise strength of \(\gamma = 0.2\), under three types of local noise: emission, pumping, and dephasing. The real and imaginary parts of the matrix elements are displayed separately for clarity. The initial GHZ state \(\rho_{\text{GHZ}}\) occupies only the corner elements \(\rho_{0,0}\), \(\rho_{0,N}\), \(\rho_{N,0}\), and \(\rho_{N,N}\).

As time progresses, the system evolves differently depending on the noise channel. Under local emission noise, excitations are lost, leading the system to gradually relax toward the ground state \(\rho_{N,N}\). For local pumping noise, excitations are injected, and the population shifts toward the fully excited state \(\rho_{0,0}\). In contrast, local dephasing noise primarily suppresses off-diagonal coherences without changing populations, driving the state toward a statistical mixture of \(\rho_{0,0}\) and \(\rho_{N,N}\).

At intermediate times \((t = t_{\mathrm{opt}})\), i.e., the time that QFI reaches its maximum, partial population spreading and coherence decay can be observed, indicating transient dynamics between the initial GHZ coherence and the mixed state. In the long-time limit \((t = 15)\), each channel leads to a distinct steady-state configuration consistent with its dissipative mechanism: \(\rho_{N,N}\) for emission, \(\rho_{0,0}\) for pumping, and \(\rho_{0,0} + \rho_{N,N}\) for dephasing. These results highlight how different local noise processes destroy entanglement in characteristic ways and determine the final structure of the density matrix.

To gain further insight into these effects, we examine the time evolution of the probe state under different noise channels, as shown in Fig.~\ref{fig10}, for $N=5$ and local noise strength $\gamma=0.2$. At $t=0$, the system is initialized in the GHZ state $\rho_0 = |\psi_{{\rm GHZ}_z}\rangle\langle\psi_{{\rm GHZ}_z}|$, which occupies the $j=N/2$ subspace of the Dicke basis with dimension $d=N+1$ and index $m=0,\ldots,d-1$. The initial density matrix contains only $\rho_{0,0}=\rho_{0,N}=\rho_{N,0}=\rho_{N,N}=0.5$, with all other elements vanishing, as shown in the first row of Fig.~\ref{fig10}. As time increases, the state population spreads across other matrix elements and, in most cases, to other $j$ blocks due to coupling with the external field and environmental noise. In the long-time limit, the state converges to $\rho_{N,N}$ under local emission, to $\rho_{0,0}$ under local pumping, and to $\rho_{0,0}+\rho_{N,N}$ under local dephasing.

\section{Quantum metrology for single-phase estimation under control Hamiltonians}
\subsection{Local noise}

To study how Hamiltonian control can mitigate local decoherence, we analyze the time dependence of the QFI $Q(t)$ under two control models: a linear drive $H_{\rm ctrl}=\chi J_x$ and a nonlinear control $H_{\rm ctrl}=\chi J_x^2$. Figure~\ref{fig11} summarizes the results for two representative local noise channels, emission and dephasing, and shows how the control strength $\chi$ modifies the metrology performance.

Figures~\ref{fig11}(a,b) correspond to local emission noise, show $Q(t)$ for different values of $\chi$ under linear (a) and nonlinear (b) control, respectively. For weak control ($\chi\lesssim 0.2$), the QFI exhibits pronounced oscillations and multiple local maxima. Increasing $\chi$ shifts the first maximum to later times, indicating an extended useful interaction window. However, for strong control the overall magnitude of $Q(t)$ is reduced, suggesting that excessive driving suppresses the accumulation of metrologically useful coherence.


Figures~\ref{fig11}(c,d) show the corresponding results for local dephasing noise. As expected, dephasing causes a faster decay of $Q(t)$ compared with emission. Nevertheless, the qualitative dependence on $\chi$ remains similar: moderate control improves $Q_{\rm max}$ and extends the interaction window, whereas strong control becomes detrimental. These results demonstrate that both linear and nonlinear control can reshape the metrology dynamics under local noise, but their optimal performance depends on the control strength and the dominant decoherence mechanism.

In the main text [Fig.~\ref{fig5}], we further evaluate the metrological gain $G(t)=Q(t)/t^2$ and its integrated value $\mathcal{I}=\int_0^T G(t)\,dt$, which are directly obtained from the QFI results shown here.

\begin{figure}[t]
\centering
    \includegraphics[width=0.6\linewidth]{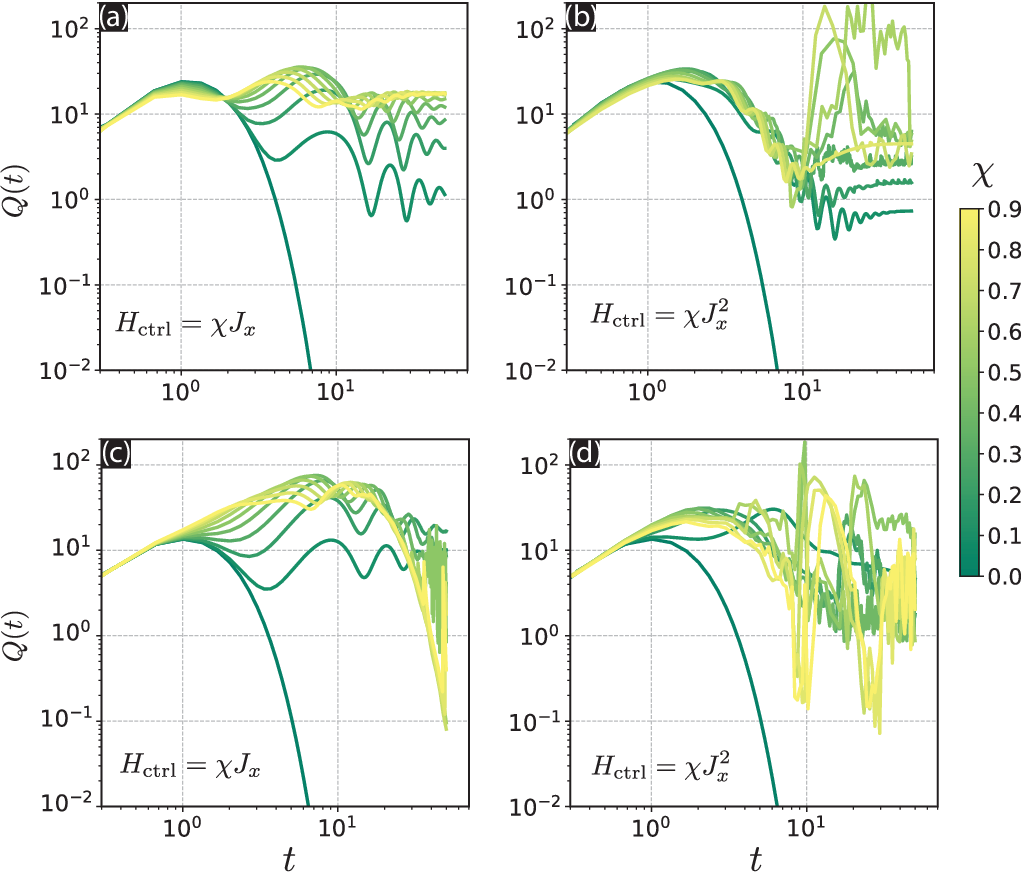}
    \caption{Quantum metrology under linear and quadratic control Hamiltonians in the presence of local noise. 
(a,b) Emission noise: QFI vs time for different values of \(\chi\), using (a) \(H_{\rm ctrl} = \chi J_x\) and (b) \(H_{\rm ctrl} = \chi J_x^2\). 
(c,d) Dephasing noise: same as (a,b). Parameters: \(N = 10\), \(\gamma_- = 0.2\) for emission noise, \(\gamma_z = 0.2\) for dephasing noise.}
    \label{fig11}
\end{figure}

\subsection{Collective noise}

We now consider the effect of Hamiltonian control in the presence of collective decoherence. Figure~\ref{fig12} summarizes the QFI dynamics under collective emission [Figs.~\ref{fig12}(a,b)] and collective dephasing [Figs.~\ref{fig12}(c,d)], using the same two control models as in the local-noise case: $H_{\rm ctrl}=\chi J_x$ and $H_{\rm ctrl}=\chi J_x^2$.

For collective emission [Figs.~\ref{fig12}(a,b)], the results show that increasing $\chi$ suppresses the peak value of $Q(t)$ and shifts the maximum toward earlier times. This behavior is opposite to the local-noise case, where moderate control can delay the optimal time.

Figures~\ref{fig12}(c,d) show the corresponding results for collective dephasing. Similar to collective emission, increasing $\chi$ generally reduces $Q_{\rm max}$ and shifts the optimal time toward earlier values. However, the degradation is more pronounced due to the stronger impact of collective dephasing on phase coherence. Overall, these results differ from the local-noise case and demonstrate that collective dissipation can significantly limit the usefulness of strong control fields. In particular, while linear and nonlinear controls can reshape the metrology dynamics, their net effect under collective noise is mainly to accelerate the evolution at the cost of reduced achievable QFI.

\begin{figure}[t]
\centering
\includegraphics[width=0.6\linewidth]{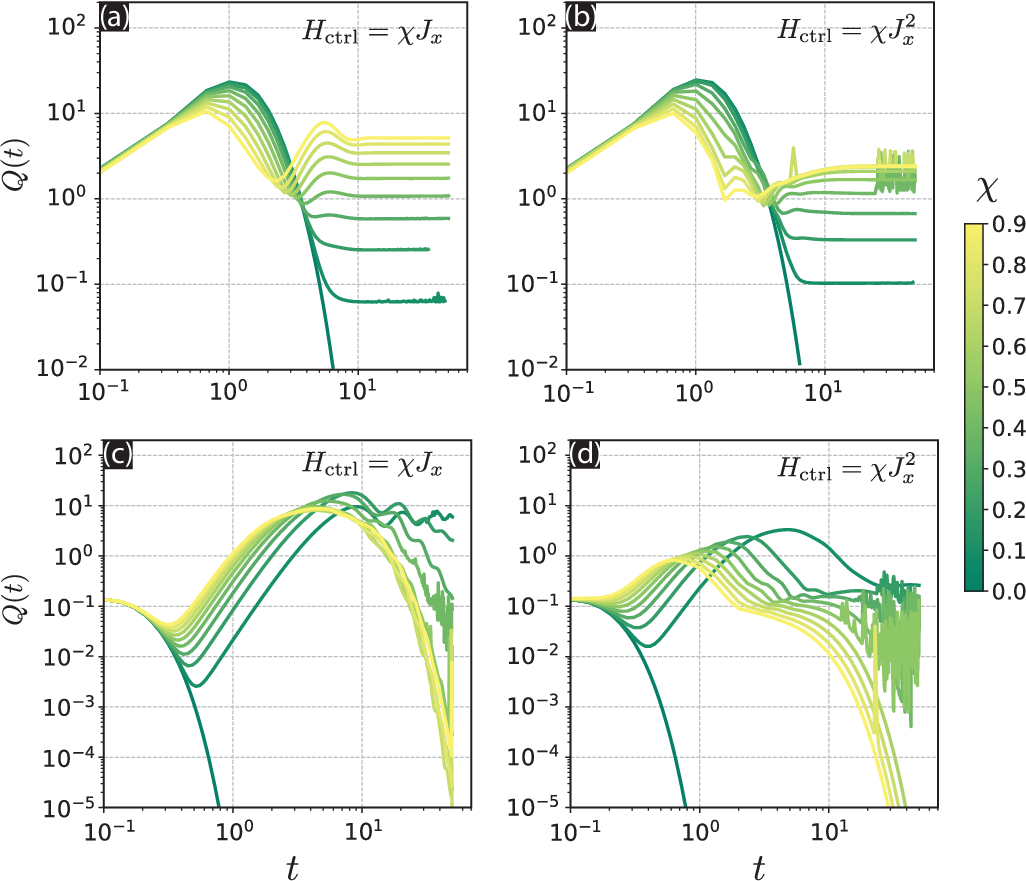}
\caption{Quantum metrology under linear and quadratic control Hamiltonians in the presence of collective noise. 
(a,b) Emission noise: QFI vs time for different values of \(\chi\), using (a) \(H_{\rm ctrl} = \chi J_x\) and (b) \(H_{\rm ctrl} = \chi J_x^2\).  
(c,d) Dephasing noise: same as (a,b). Parameters: \(N = 10\), \(\Gamma_- = 0.2\) for emission noise, \(\Gamma_z = 0.2\) for dephasing noise.}
\label{fig12}
\end{figure}

In the main text [Fig.~\ref{fig6}], we further evaluate the metrological gain $G(t)=Q(t)/t^2$ and its integrated value $\mathcal{I}=\int_0^T G(t)\,dt$, which are directly obtained from the QFI results shown here.

\end{widetext}

\bibliography{refs}

\end{document}